\documentclass[pra, showpacs, twocolumn, a4paper, nofootinbib]{revtex4}
\usepackage{graphicx}
\usepackage{amsmath, amsfonts, amssymb, bm}
\usepackage{amsmath}
\usepackage{verbatim}
\usepackage{bm}
\usepackage{graphicx,psfrag,bm,times,amsmath,amssymb}

\begin{document}

\title{Squeezed single-atom laser in a photonic crystal}

\author{Rong \surname{Tan}$^{a}$}
\author{Gao-xiang \surname{Li}$^{a}$}
\email{gaox@phy.ccnu.edu.cn}
\author{Zbigniew \surname{Ficek}$^{b}$}
\affiliation{$^{a}$Department of Physics, Huazhong Normal University, Wuhan 430079, China\\
$^{b}$Department of Physics, School of Physical Sciences, The University of Queensland,
Brisbane, Australia 4072}

\date{\today}

\begin{abstract}
We study non-classical and spectral properties of a strongly driven single-atom laser  engineered
within a photonic crystal that facilitates a frequency-dependent reservoir. In these studies,
we apply a dressed atom model approach to derive the master equation of the system and study the properties  of the dressed laser under the frequency dependent  transition rates.
By going beyond the secular approximation in the dressed-atom cavity-field interaction, we find that
if, in addition, the non-secular terms are included into the dynamics of the system, then non-linear processes can occur that lead to interesting new aspects of cavity field behavior.
We calculate variances of the quadrature phase amplitudes and the incoherent part of the spectrum of the cavity field and show that they differ qualitatively from those observed under the secular approximation. In particular, it is found that the non-linear processes lead to squeezing of the fluctuations of the cavity field below the quantum
shot noise limit. The squeezing depends on the relative population of the dressed states of the system
and is found only if there is no population inversion between the dressed states.
Furthermore, we find a linewidth narrowing below the quantum limit in the spectrum of the cavity field that is achieved only when the secular approximation is not made. An interpretation of the linewidth narrowing is provided in terms of
two phase dependent noise (squeezing) spectra that make up the incoherent spectrum. We establish that the linewidth narrowing is due squeezing of the fluctuations in one quadrature phase components of the cavity~field.
\end{abstract}
\pacs{42.55.Tv, 42.50.Ct, 42.50.Ar} \maketitle

\section{Introduction}
\ \ A single-atom laser, in which no more than one atom is present
in an optical resonator, is one of the key systems to study quantum
effects in the interaction of electromagnetic fields with
matter~\cite{zl91,lz90,lg97,lc07,kk08}. Although a single-atom laser
is admittedly an elementary model, it has the advantage over a
multi-atom laser that in any practical realization of the laser one
is not concerned with many difficulties such as fluctuations of the
number of atoms. With the present trapping and cooling techniques,
the atom can easily localized in a small region within the
Lamb-Dicke regime. Because of its simplicity, many quantum features
of single-atom lasers, such as sub-Poissonian photon statistics,
photon antibunching, squeezing and vacuum Rabi splitting have been
predicted and experimentally observed~\cite{mck,tp}.

\ \ The primary obstacle in the realization of one-atom lasers is
spontaneous emission.  The reason is that spontaneous emission is a
source of noise that leads to emission of photons into modes
different from the cavity mode. Therefore, it is not surprised that
many schemes have been proposed to reduce spontaneous emission. It
has been demonstrated that spontaneous emission can be reduced if an
atom is located in a squeezed vacuum~\cite{df04}, or inside a cavity
that facilitates an exclusive spatial selection of radiation modes
coupled to the atom~\cite{p46,k81,gr83,hc87}. The cavity appears as
a frequency filter for spontaneous emission. A further reduction of
spontaneous emission can be achieved through a dynamical means that
unlike the conventional method of coupling an atom into the cavity
mode, one could first drive the atom with a strong laser field and
then couple the resulting dressed-atom system with the cavity
mode~\cite{lm87,lm88}. When the Rabi frequency of the driving laser
is much larger than the cavity bandwidth, one can tune the cavity
mode to one of the dressed-atom transition frequencies thereby
eliminating spontaneous emission on the other transitions. Within
this approach, a wide variety of quantum and spectral features, such
as atomic population inversion~\cite{s88,qf93,zs98}, and dynamical
suppression and narrowing of the spectral lines~\cite{fq94,n90} has
been predicted and some of them verified
experimentally~\cite{zlm88,lz89,gz91,lw93}.

\ \ Recently, Florescu {\it et al.}~\cite{lf} have applied the
dressed-atom approach to a single-atom laser that incorporates the
dynamical suppression of spontaneous emission inside a photonic
crystal. They have applied the secular approximation in the coupling
of a dressed two-level atom to the cavity mode that the Rabi
frequency of the dressing field is much larger than the coupling
constant between the atom and the cavity field. By appropriate
tuning of the cavity mode and the dressed-atom transition
frequencies, they have shown that the variances of the cavity field
amplitude and the linewidth of the cavity field spectrum can be
reduced to the quantum shot noise limit. These effects result from
the filtering property of the photonic crystal that effectively
forms a frequency dependent reservoir of a specific spectral
function (step function) of the radiation modes. A photonic crystal
that is a periodic dielectric structure, can prohibit light
propagation over a continuous range of frequencies, irrespective of
the direction of propagation~\cite{p1,p2}. It is also known that in
photonic crystals extremely small micro-cavity mode volumes and very
high cavity $Q$ factors  can be realized~\cite{gr,op,sj}.  It is
associated with the unique properties of photonic crystals, i.e.,
the photonic density of states within or near a photonic band gap
can almost vanish or exhibit discontinuous changes as a function of
frequency with appropriate engineering, which is essentially
different from its free-space counterpart.

\ \ Although the secular approximation is often entirely adequate to
describe the interaction of the cavity field with a dressed-atom
system, there are circumstances when the non-secular terms might be
important, for example in the strong coupling limit of the cavity
field to the dressed-atom system. One could argue that even under
these circumstances, the non-secular terms would only make small
corrections to the dynamics of the system. In the following,  we
show that this is not the case, they may lead to new aspects of
cavity field behavior that differ qualitatively from those observed
under the secular approximation. To show this, the dressed-atom
approach for a driven two-level atom placed inside a photonic
crystal is generalized to deal with non-secular terms in the
atom-cavity field interaction. The method used in this paper is
similar to that used by Florescu {\it et al.}~\cite{lf} but with one
significant difference, we do not make the secular approximation in
the coupling of the cavity mode to the dressed-atom system. Instead,
we consider the role of the non-secular terms in the operation of a
strongly driven single-atom laser. We present solutions for the
steady-state variance of the quadrature components of the cavity
field amplitude and the incoherent part of the spectrum of the
cavity field. In the course of the calculations, we observe that the
inclusion of the non-secular terms results in a {\it double} coupling of
the cavity field to the dressed atom and leads to non-linear terms
in the master equation of the system that reflect a possibility of
non-linear processes between the dressed states when the secular
approximation is not made. Our results demonstrate that these
non-linear processes lead to the reduction of the fluctuations of
the cavity field amplitude below the quantum shot noise limit. In
addition, we find that the spectral line of the cavity field may now
be narrowed below the quantum limit giving a subnatural linewidth of
the emitted field. A qualitative understanding of the origin of the
line narrowing is obtained by writing the incoherent part of the
cavity field spectrum in terms of the phase dependent noise
(squeezing) spectra. We show that the narrowing results from
negative values of the squeezing spectrum of one quadrature phase
component of the field.

\ \ The paper is organized as follows. In Sec. II, we present the
model Hamiltonian and derive the master equation for a strongly
driven two-level atom coupled to a single-mode cavity and placed
inside a photonic band gap material that acts as a frequency
dependent reservoir.  In Sec. III, we derive the basic equations of
motion for the expectation values of the atomic and field
correlation functions are find their steady-state values. We then
apply them in Sec.~IV to investigate variances of the quadrature
phase components of the cavity field that determine the fluctuations
of the cavity field amplitude, and in Sec.~V to the calculations of
the spectrum of the cavity field. We calculate the incoherent part
of the spectrum and then analyze its properties in terms of the
squeezing spectra. Finally, we summarize the results in Sec. VI.

\section{Master Equation}\label{sec2}

\ \ We consider a single two-level atom with ground state
$|1\rangle$ and excited state $|2\rangle$ separated by the
transition frequency $\omega_{a}$. The atom is coupled to a single
mode of a high-$Q$ microcavity engineered within a photonic crystal
with coupling constant $g$, and is driven by a coherent external
laser field of a frequency $\omega_L$ and the resonant Rabi
frequency $\epsilon$. In addition, the atom is damped at the rate
$\gamma$ by spontaneous emission to modes other than the cavity
mode. In practice this model can be realized by embedding a quantum
dot in a dielectric microcavity (defect) placed within a two-mode
waveguide channel in a 2D PBG microchip \cite{lf}. In a photonic
band gap material, one mode of the waveguide channel is engineered
to produce a large discontinuity in the local photon density of
states near the atom, and another mode is used to propagate the pump
beam. By suitable engineering, it is possible to realize a strong
coupling of the quantum dot to both the pumping waveguide mode and
the high-$Q$ cavity mode \cite{lf}. For simplicity, we treat the
driving external field classically and work in the interaction
picture. The total system is described by a Hamiltonian that under
the electric-diople and the rotating-wave approximations can be
written as
\begin{eqnarray}
H = H_{0}+H_{1} ,\label{e1}
\end{eqnarray}
where the first term
\begin{eqnarray}
H_{0} &=& \hbar\Delta_{c}a^{\dag}a+\frac{1}{2}\hbar\Delta_{a}\sigma_{3} \nonumber \\
&& +\hbar\epsilon(\sigma_{12}+\sigma_{21})
+\hbar\sum_{\lambda}\Delta_{\lambda}a_{\lambda}^{\dag}a_{\lambda} \label{e2}
\end{eqnarray}
is the noninteracting Hamiltonian of the driven atom plus the cavity mode plus the photonic crystal radiation
reservoir modes, and the second term
\begin{eqnarray}
H_{1} &=& i\hbar{g}(a^{\dag}\sigma_{12}-\sigma_{21}a) \nonumber \\
&& +i\hbar\sum_{\lambda}g_{\lambda}(\omega_{\lambda})
(a^{\dag}_{\lambda}\sigma_{12}-\sigma_{21}a_{\lambda})  \label{e3}
\end{eqnarray}
is the interaction Hamiltonian between the atom and  the cavity mode and the photonic crystal
vacuum radiation modes.

Here, $a$ and $a^{\dag}$ are the cavity-mode annihilation and creation operators, $a_{\lambda}$
and $a^{\dag}_{\lambda}$ are the photonic crystal radiation reservoir annihilation and creation
operators, $\sigma_{ij}$ are the bare atomic operators, $\sigma_{ij}=|i\rangle\langle{j}|\ (i,j=1,2)$, and
$\sigma_{3}=\sigma_{22}-\sigma_{11}$ describes the bare atomic inversion.
The parameter $\Delta_{a}=\omega_{a}-\omega_{L}$ denotes the detuning of
the atomic resonance frequency $\omega_{a}$ from the frequency $\omega_{L}$ of the driving
laser field, $\Delta_{c}=\omega_{c}-\omega_{L}$ is the detuning of the cavity frequency from the
frequency of the laser field, and $\Delta_{\lambda}=\omega_{\lambda}-\omega_{L}$ is
the detuning  of the reservoir frequency $\omega_{\lambda}$ of a mode $\lambda$ from the laser frequency.

\ \ The coefficient $g$ describes the strength of the coupling
between the atom and the cavity mode that we assume to be constant
independent of frequency,  and $g_{\lambda}(\omega_{\lambda})$
describes the strength of the coupling between the atom and the
vacuum  modes of the photonic crystal. It contains the information
about the frequency dependent mode structure of the photonic crystal
and can be written as
\begin{eqnarray}
g_{\lambda}(\omega_{\lambda}) = g_{\lambda}D(\omega_{\lambda}) ,
\end{eqnarray}
where $g_{\lambda}$ is a constant proportional to the dipole moment of the atom, and
$D(\omega_{\lambda})$ is the transfer function of the photonic crystal, the absolute value square
of which can be identified as the Airy function of a frequency dependent radiation
reservoir~\cite{Yar}.

\ \ The explicit form of $D(\omega_{\lambda})$ depends on the type
of radiation reservoir used. For a photonic band gap material, the
transfer function is in the form of the unit step function
$|D(\omega_{\lambda})|^{2} = u(\omega_{\lambda}-\omega_{b})$, where
$\omega_{b}$ is the photonic density of states band edge frequency.
Thus, $|D(\omega_{\lambda})|^{2} =0$ for
$\omega_{\lambda}<\omega_{b}$ and $|D(\omega_{\lambda})|^{2} =1$ for
$\omega_{\lambda}>\omega_{b}$. Since the frequencies
$\omega_{\lambda}<\omega_{b}$ are forbidden in the band gap
material, it is possible to selectively eliminate spontaneous
emission at some frequencies of the driven atom.

\ \ The strong driving field can be viewed as a dressing field for
the atom. Therefore, we begin by diagonalizing the atomic part of
the Hamiltonian together with the interaction of the atom with the
laser field
\begin{eqnarray}
H_{af} = \frac{1}{2}\hbar\Delta_{a}\sigma_{3} +\hbar\epsilon(\sigma_{12}+\sigma_{21}) , \label{af}
\end{eqnarray}
to find the eigenstates (dressed states) of the combined atom plus driving field system.
Since the driving field is treated classically in our calculations, we find the so-called semiclassical
dressed states
\begin{eqnarray}
|\tilde{1}\rangle &=& \cos{\phi}|1\rangle+\sin{\phi}|2\rangle,\nonumber \\
|\tilde{2}\rangle &=& \sin{\phi}|1\rangle-\cos{\phi}|2\rangle ,\label{ds}
\end{eqnarray}
where $\cos^{2}{\phi} = (1+\Delta_{a}/\Omega)/2$ with the angle
$\phi$ defined such that $0\leq \phi \leq \pi/2$. The dressed states
form non-degenerate doublets that are separated in energy by $\hbar\omega_{L}$, and
the states of the doublet are split by $\hbar 2\Omega$, where
$\Omega=(4\epsilon^{2}+\Delta_{a}^{2})^{1/2}$ is the Rabi frequency
of the detuned field.

We now couple the dressed states to the cavity field and to the photonic crystal vacuum modes.
First, we replace the atomic operators by the
dressed-state operators
\begin{eqnarray}
\sigma_{12} &=& -\frac{1}{2}\sin(2\phi)R_{3} + \sin^{2}\phi R_{21} -\cos^{2}\phi R_{12} ,\nonumber \\
\sigma_{21} &=& -\frac{1}{2}\sin(2\phi)R_{3} + \sin^{2}\phi R_{12} -\cos^{2}\phi R_{21} , \nonumber \\
\sigma_{3} &=& -\cos(2\phi) R_{3} +\sin(2\phi)\left(R_{12}+R_{21}\right) ,
\end{eqnarray}
where $R_{ij}= |\tilde{i}\rangle \langle\tilde{j}|$ are the dressed-atom dipole operators and
$R_{3} = R_{22}-R_{11}$.
Next, we perform  the unitary "dressing" transformation of the interaction Hamiltonian
\begin{eqnarray}
\tilde{H}_{1}= \exp(i\tilde{H}_{0}t)H_{1}\exp(-i\tilde{H}_{0}t) ,
\end{eqnarray}
with
\begin{eqnarray}
\tilde{H}_{0} = \Omega R_{3} + \Delta_{c}a^{\dag}a
+ \sum_{\lambda}\Delta_{\lambda}a_{\lambda}^{\dag}a_{\lambda} ,
\end{eqnarray}
and obtain the interaction Hamiltonian between the dressed atom and both the cavity mode and the
photonic crystal vacuum modes
\begin{eqnarray}
\tilde{H}_{1} &=& i\hbar{g}\left(sca^{\dag}R_{3}e^{i\Delta_{c}t}
+c^{2}a^{\dag}R_{12}e^{i(\Delta_{c}-2\Omega)t}\right. \nonumber \\
&&\left. -s^{2}a^{\dag}R_{21}e^{i(\Delta_{c}+2\Omega)t} - \rm{H.c.}\right) \nonumber\\
&& +i\hbar\sum_{\lambda}g_{\lambda}\left(sca_{\lambda}^{\dag}R_{3}e^{i\Delta_{\lambda}t}
+c^{2}a_{\lambda}^{\dag}R_{12}e^{i(\Delta_{\lambda}-2\Omega)t}\right. \nonumber \\
&&\left. -s^{2}a_{\lambda}^{\dag}R_{21}e^{i(\Delta_{\lambda}+2\Omega)t} -\rm{H.c.}\right) ,\label{e9}
\end{eqnarray}
where $s=\sin\phi$ and $c=\cos\phi$.

\ \ The Hamiltonian (\ref{e9}) describes the interaction of the
dressed atom with the cavity field and with the vacuum modes. We see
that in the dressed-atom picture, the cavity frequency and the
vacuum modes are tuned to the dressed-state transitions that occur
at three characteristic frequencies, $\Delta_{c}$ and $\Delta_{c}\pm
2\Omega$. By matching the cavity field frequency to one of the
dressed states frequencies, we may manipulate the strange of the
interaction between the driven system and the cavity mode.

\ \ Our aim is to derive from the Hamiltonian (\ref{e9}) the master
equation for a reduced density operator of the driven atom and the
cavity field. It is obtained by tracing the density operator of the
total system over the photonic crystal radiation reservoir
variables. On carrying out this procedure, it is found that in the
dissipative part of the master equation certain terms are slowly
varying in time while the others are oscillating with frequencies
$2\Omega$ and $4\Omega$. Since we are interested in the case where
the Rabi frequency $\Omega$ is much larger than the atomic and
cavity damping rates
\begin{eqnarray}
\Omega \gg \gamma, \kappa ,
\end{eqnarray}
\ \ we can invoke the secular approximation that consists of
dropping these rapidly oscillating terms. These terms, if kept in
the master equation, would make corrections to the  dynamics of the
system of the order of $\gamma/\Omega$, and thus completely
negligible. We therefore find that after discarding the rapidly
oscillating terms in the dissipative part of the master equation,
the time evolution of the reduced density operator is of the form
\begin{align}
\frac{\partial{\rho}}{\partial{t}} & =  gsc[a^{\dag}R_{3}e^{i\Delta_{c}t}
-aR_{3}e^{-i\Delta_{c}t},\rho]\nonumber\\
&+gc^{2}[a^{\dag}R_{12}e^{i(\Delta_{c}-2\Omega)t}-aR_{21}e^{-i(\Delta_{c}-2\Omega)t},\rho]\nonumber\\
&-gs^{2}[a^{\dag}R_{21}e^{i(\Delta_{c}+2\Omega)t}
-aR_{12}e^{-i(\Delta_{c}+2\Omega)t},\rho]  \nonumber \\
& +{\cal L}_{a}{\rho}+ {\cal L}_{c}{\rho} ,\label{me}
\end{align}
where
\begin{align*}
{\cal L}_{a}{\rho}&=\frac{1}{2}\gamma_{0}\left(R_3\rho{R_3}-\rho\right)
+\frac{1}{2}\gamma_{-}\left(R_{21}\rho{R_{12}}-R_{12}R_{21}\rho\right) \nonumber\\
&+\frac{1}{2}\gamma_{+}\left(R_{12}\rho{R_{21}}-R_{21}R_{12}\rho\right) +\rm{H.c.} \nonumber\\
{\cal L}_{c}{\rho}&= \frac{1}{2}\kappa \left(2a\rho{a}^\dag-{a}^\dag{a}\rho-\rho{a}^\dag{a}\right) ,
\end{align*}
describe spontaneous dynamics between the dressed states of the system and of the cavity mode,
respectively. The parameters
\begin{eqnarray}
\gamma_{0} &=& s^{2}c^{2}\gamma |D(\omega_{L})|^{2} +(c^{2} -s^{2})\gamma_{p} ,\nonumber \\
\gamma_{-} &=& s^{4}\gamma |D(\omega_{L}-2\Omega)|^{2} + 4s^{2}c^{2}\gamma_{p} ,\nonumber \\
\gamma_{+} &=& c^{4}\gamma |D(\omega_{L}+2\Omega)|^{2}+4s^{2}c^{2}\gamma_{p}
\end{eqnarray}
determine the damping rates between the dressed states of the system. They also include the contribution of the dephasing rate~$\gamma_{p}$ which may arise from scattering of phonons of the host crystal on the atom embedded in the solid part of the dielectric material.
The coefficient $\gamma_{0}$ corresponds to spontaneous emission occurring at two transitions
of the dressed atom; One from the lower dressed state $|\tilde{1}\rangle $ to the lower  dressed state
of the manifold below and the other from the upper dressed state $|\tilde{2}\rangle $ to the upper dressed state of the manifold below. These transitions occur at frequency $\omega_{L}$.
The coefficient $\gamma_{+}$ corresponds to spontaneous emission from the upper dressed
state to the lower  dressed state of the manifold below and occurs at frequency $\omega_{L}+2\Omega$,
whereas the coefficient $\gamma_{-}$ corresponds to spontaneous emission from the lower dressed
state to the upper dressed state of the manifold below and occurs at frequency $\omega_{L}-2\Omega$.

\ \ An essential feature of the master equation (\ref{me}) is the
presence of the oscillatory terms in the coherent part of the
evolution involving the dressed atom and the cavity field. In
experimental practice it might imply that there is no restriction on
the relation between the Rabi frequency $\Omega$ and the coupling
constant $g$, and the master equation can be applied to both
situations, where the coupling constant $g$ is much smaller or
comparable to the Rabi frequency.

\ \ As we have already mentioned, we work in the strong-coupling
regime of $\Omega \gg g\gg \gamma,\kappa$. The latter inequality has been invoked into Eq.~(\ref{me}) 
to neglect the effect of non-secular terms in the dissipative part of the master equation, but to keep 
the non-secular terms in the interaction of the cavity field with the dressed atom. This condition is satisfied in a typical band-gap material, where strong couplings $g$ and significant reductions of the 
spontaneous emission rate $\gamma$ are easily achieved due to the confinement of the atom to 
an extremely small volume with a significantly suppressed density of the vacuum 
modes. For example, given the electric dipole moment of an atom of 
$\mu\approx 10^{-29}$C m$^{-1}$ and the cavity mode volume of $V\approx 10^{-6}$\ m$^{3}$, then the atom cavity coupling will be $g\approx 10^{10}$~Hz~\cite{lf,gr}.
With the spontaneous emission rate of the atom in a free space of $\gamma \approx 10^{8}$\ Hz, the inequality $g\gg \gamma$ is easy satisfied. The parameter $\kappa$, that describes the damping of the cavity field, is equal to zero in an ideal cavity. However, in a realistic structure $\kappa \neq 0$, that   
arises from a weak coupling of the cavity mode to the waveguide mode of the band-gap material. This coupling may result from defects or disorders in the waveguide channel caused by the manufacturing process that can be kept at a minimal level~\cite{lf,p2,gr,sj}.
Effectively, the waveguide channel acts as an external reservoir of vacuum modes to which photons can escape from the cavity mode. Thus, the inequality $g\gg \gamma,\kappa$ is realistic and should be easy to satisfy within the current band-gap material technology.  

\ \ In the following, we focus on the case of
$\Delta_{c}\approx 0$, i.e., the cavity field tuned close to the
central frequency of the dressed atom. In this limit, the master
equation (\ref{me}) contains terms that are time independent and
thus corresponding to the resonant interaction of the cavity field
with the dressed atom. It also contains terms that have an explicit
time dependence of the form $\exp(\pm 2i\Omega t)$. These terms
correspond to a dispersive (non-resonant) interaction of the cavity
field with the dressed atom, which induces interesting new property
as we will show below.

\ \ We first perform a canonical transformation of the master
equation (\ref{me}) by
\begin{eqnarray}
\tilde{\rho} = \exp(i\Delta_{c}a^{\dagger}at)\rho\exp(-i\Delta_{c}a^{\dagger}at) ,
\end{eqnarray}
and find that the master equation for the transformed density operator takes the form
\begin{eqnarray}
\frac{\partial{\tilde{\rho}}}{\partial{t}}  =  -i\left[H_{0}^{\prime} +H^{\prime}(t),\tilde{\rho}\right]
 +{\cal L}_{a}{\tilde{\rho}}+ {\cal L}_{c}{\tilde{\rho}} ,\label{met}
\end{eqnarray}
where
\begin{eqnarray}
H_{0}^{\prime} &=& gsc\left(a^{\dag}R_{3} -aR_{3}\right) +\Delta_{c}a^{\dagger}a ,\nonumber\\
H^{\prime}(t) &=& ig\left[\left(c^{2}a^{\dag}+s^{2}a\right)R_{12}e^{-i2\Omega{t}} -{\rm H.c.}\right] .
\end{eqnarray}
Note that the Hamiltonian $H^{\prime}(t)$ disappears under the
secular approximation on the interaction of the dressed-atom with
the cavity field. Due to the presence of the rapidly oscillating
terms, the Hamiltonian $H^{\prime}(t)$ can be treated as a perturber
to the time-independent Hamiltonian $H_{0}^{\prime}$. Since
$g\ll\Omega$, we can perform a second-order perturbation
calculations with respect to $g$ and find an effective Hamiltonian
\cite{method} that can be written as
\begin{eqnarray}
H'_{\rm{eff}} &=&  -iH'(t)\int{H'(t')dt'} =  \frac{\hbar{g^2}}{2\Omega}
\left[s^2c^2 R_{3}\left(a^2 +{a}^{\dag2}\right)\right. \nonumber\\
&& \left. + (s^4+c^4)\left(R_3a^\dag{a}+\frac{1}{2}R_3\right)\right] .
\end{eqnarray}
In this case, the master equation (\ref{me}) takes the form
\begin{align}
\frac{\partial{\tilde{\rho}}}{\partial{t}} &= -i\left[\left(\Delta_{c}
+\frac{{g^{2}_{2}}}{2\Omega}R_{3}\right) {a}^\dag{a},\tilde{\rho}\right]
-i\frac{{g^{2}_{2}}}{4\Omega}\left[R_3,\tilde{\rho}\right] \nonumber\\
&+g_{1}[({a}^\dag-a)R_3,\tilde{\rho}]
-i\frac{{g_1^2}}{2\Omega}\left[R_{3}(a^2 +{a}^{\dag2}),\tilde{\rho}\right]  \nonumber\\
&+{\cal L}_{a}\tilde{\rho} + {\cal L}_{c}\tilde{\rho} ,\label{ma1}
\end{align}
where
\begin{equation}
g_{1} = gsc \quad {\rm and}\quad g_{2} = g\sqrt{(s^4+c^4)}
\end{equation}
are  the "effective" coupling constants of the cavity field to the dressed-atom system.

\ \ The important new feature of the master equation (\ref{ma1}) is
in the presence of an additional non-linear terms that results in an
effective {\it double} coupling of the cavity field to the dressed
atom. The first term on the right-hand side of Eq.~(\ref{ma1})
represents a shift of the cavity frequency. It includes a
dressed-atom inversion dependent shift that arises solely from the
non-secular terms in the dressed-atom picture. The second term
represents a shift of the dressed-atom frequencies that also arises
from the presence of the non-secular terms. The third term
represents a resonant coupling of the cavity field to the dressed
atom at frequency $\omega_{c}\approx \omega_{L}$, and finally the
fourth term describes a dispersive coupling through the off-resonant
Rabi sideband frequencies and is determined by non-linear two-photon
absorption and emission processes between the dressed states.
Thus, the presence of the non-secular terms results in the non-linear coupling of the cavity field
to the dressed atom and shifts of the cavity and dressed-atom resonance frequencies.
The amplitude of these contributions is of order $g^{2}/\Omega$ and therefore is small, but not
necessarily too small to make detectable contributions to the dynamics of the system.
In what follows, we shall show that the non-linear terms in the master equation (\ref{ma1})
play in fact an important role in the dynamics of the system and will study in details non-classical
and spectral properties of the out-put field of the laser operating with a resonator engineered
in a photonic band gap material.

\section{Equations of motion}

\ \ In this section we investigate the influence of the non-secular
terms on the properties of a driven single-atom laser. The
properties can all be expressed in terms of single-time and two-time
expectation values of the dressed-atom and the cavity field
operators. The master equation~(\ref{ma1}) enables us to derive
equations of motion for expectation values of an arbitrary
combination of the atomic and cavity field operators. In particular,
for the dressed-atom population inversion and the cavity field
amplitudes, we find the following closed set of equations of motions
\begin{align}
\frac{d}{dt}\langle{R_{3}}\rangle =& -\gamma_{2}-\gamma_{1}\langle{R_{3}}\rangle, \nonumber\\
\frac{d}{dt}\langle{a}\rangle =& -\left(\frac{1}{2}\kappa -i\Delta_{c}\right)\langle{a}\rangle +g_{1}\langle{R_{3}}\rangle\nonumber\\
&-i\frac{g^{2}_{2}}{2\Omega}\langle{R_{3}a}\rangle-i\frac{g_{1}^{2}}{\Omega}\langle{R_{3}a^{\dag}}\rangle,\nonumber\\
\frac{d}{dt}\langle{R_{3}a}\rangle =& \  g_{1} - \left(\gamma_{1}+\frac{1}{2}\kappa - i\Delta_{c}\right)\langle{R_{3}a}\rangle  \nonumber\\
&-\gamma_{2} \langle{a}\rangle - i\frac{g^{2}_{2}}{2\Omega}\langle{a}\rangle -i\frac{g_{1}^{2}}{\Omega}\langle{a^{\dag}}\rangle,
\label{e19}
\end{align}
and equations of motion for $\langle a^{\dagger}\rangle$ and $\langle{R_{3}a^{\dagger}}\rangle$ are obtained by Hermitian conjugate of the above equations.

\ \ For the expectation values involving higher order combinations
of the operators, such us the number of photons, the master equation
leads to the following set of equations of motion
\begin{align}
\frac{d}{dt}\langle{a^{\dag}a}\rangle =& -\kappa\langle{a^{\dag}a}\rangle
+g_{1}\left(\langle{R_{3}a^{\dag}}\rangle+ \langle{R_{3}a}\rangle\right) \nonumber\\
&-i\frac{g_{1}^{2}}{\Omega}\left(\langle{R_{3}a^{\dag2}}\rangle-\langle{R_{3}a^{2}}\rangle\right),
\nonumber \\
\frac{d}{dt}\langle{a^{\dag2}}\rangle =&
i\frac{g_1^{2}}{\Omega}\langle{R_{3}}\rangle -(\kappa
+2i\Delta_{c})\langle{a^{\dag2}}\rangle
+2g_{1}\langle{R_{3}a^{\dag}}\rangle \nonumber \\
&+i\frac{g^{2}_{2}}{\Omega}\langle{R_{3}a^{\dag2}}\rangle
+2i\frac{g_1^{2}}{\Omega}\langle{R_{3}a^{\dag}a}\rangle ,\nonumber \\
\frac{d}{dt}\langle{R_{3}a^{\dag2}}\rangle =& \ i\frac{g_{1}^{2}}{\Omega}
 -\left(\gamma_{1} +\kappa +2i\Delta_{c}\right)\langle{R_{3}a^{\dag2}}\rangle
+ 2g_{1}\langle{a^{\dag}}\rangle \nonumber \\
& -\gamma_{2}\langle{a^{\dag2}}\rangle
+i\frac{g^{2}_{2}}{\Omega}\langle{a^{\dag2}}\rangle
+2i\frac{g_{1}^{2}}{\Omega}\langle{a^{\dag}a}\rangle ,\nonumber \\
\frac{d}{dt}\langle{R_{3}a^{\dag}a}\rangle =&
-(\gamma_{1}+\kappa)\langle{R_{3}a^{\dag}a}\rangle
+g_{1}(\langle{a^{\dag}}\rangle+\langle{a}\rangle) \nonumber \\
&-\gamma_{2}\langle{a^{\dag}a}\rangle
-i\frac{g_{1}^{2}}{\Omega}(\langle{a^{\dag2}}\rangle-\langle{a^{2}}\rangle) ,\label{e20}
\end{align}
and equations of motion for $\langle{a^{2}}\rangle$ and  $\langle{R_{3}a^{2}}\rangle$ are obtained by the Hermitian conjugate of the equations of motion for $\langle{a^{\dag2}}\rangle$ and
$\langle{R_{3}a^{\dag2}}\rangle$, respectively.

\ \ The quantities $\gamma_{1}$ and $\gamma_{2}$ occurring in
Eqs~(\ref{e19}) and (\ref{e20}) are given in terms of the damping
rates $\gamma_{+}$ and $\gamma_{-}$, and are defined as
\begin{eqnarray}
\gamma_{1}&=& \gamma_{+}c^{4}+\gamma_{-}s^{4}+8\gamma_{p}s^{2}c^{2}, \nonumber \\
\gamma_{2}&=& \gamma_{+}c^{4}-\gamma_{-}s^{4}. \label{e21a}
\end{eqnarray}
We see that the quantities $\gamma_{1}$ and $\gamma_{2}$ depend only on the transition rates centered at the dressed states resonances corresponding to the Rabi sidebands.
Moreover, the quantity $\gamma_{2}$ can be negative, whereas $\gamma_{1}$ is positive for all
values of the parameters involved. In addition, $\gamma_{1}\geq \gamma_{2}$ even in the absence
of the dephasing process, $\gamma_{p}=0$.

\section{Steady-state solutions}

\ \ We proceed here to discuss the steady-state properties of the correlation functions for
 $\Delta_{c}=0$ that are listed in Appendix I.
 In this simplified case, we obtain solutions in a physically transparent form that
 will allow one to gain physical insight into how the non-secular terms and structured band gap
material can modify the properties of the dressed-atom system and
the cavity field. The exact steady-state solutions of
Eqs.~(\ref{e19}) and (\ref{e20}), valid for an arbitrary detuning
$\Delta_{c}$ are quite lengthy and will be computed numerically in the next section.

\ \ Before commencing our analysis of the properties of the cavity
field, we first look into properties of the dressed atom. We observe
that the equation of motion for the expectation value of the atomic
operator $R_{3}$ is decoupled from the remaining equations. Hence,
it has a simple steady-state solution
\begin{eqnarray}
\langle R_{3}\rangle = -\frac{\gamma_{2}}{\gamma_{1}} .\label{e21}
\end{eqnarray}
This shows that the steady-state value of the dressed inversion operator is not affected by the
non-secular terms. However, it depends crucially on the quantities $\gamma_{1}$ and
$\gamma_{2}$ that, on the other hand, depend on the mode structure of the band gap material.
The inversion varies between $+1$ and $-1$ corresponding to locking the population
in the dressed state $|\tilde{2}\rangle$ and $|\tilde{1}\rangle$, respectively. According to
Eq.~(\ref{e21a}), it happens when there is no spontaneous emission on one of the Rabi frequency
of the dressed system.

\ \ Let us now examine the stationary properties of the cavity
field. In the steady-state regime and for the case of the cavity
frequency and the atomic transition frequency tuned on resonance
with the central frequency of the dressed-atom system,
$\Delta_{c}=0$ and $\Delta_{a}=0$, the expectation values of the
cavity field operators can be expressed as
\addtocounter{equation}{1}
\begin{align}
\langle{a}\rangle&= -\frac{2g_{1}\gamma_{2}}{\gamma_{1}\kappa}\left[
1 + i\frac{4g_{1}^{2}\gamma_{1}} {\kappa\gamma_{2}\Omega}
-i\frac{8g_{1}^{2}\left(\gamma_{1}^{2}-\gamma_{2}^{2}\right)}
{\kappa\gamma_{2}\left(2\gamma_{1}+\kappa\right)\Omega}\right] ,\nonumber \\
\langle{a^{2}}\rangle&= \frac{4g_{1}^{2}}{\kappa^{2}}\left[ 1 -
\frac{2\left(\gamma_{1}^{2}-\gamma_{2}^{2}\right)}
{\gamma_{1}\left(2\gamma_{1}+\kappa\right)}\right] \nonumber \\
&+ i\frac{g_{1}^{2}\gamma_{2}}{\kappa\gamma_{1}\Omega}\left\{1
+\frac{32g_{1}^{2}}{\kappa^{2}}\left[ 1- \frac{(3\kappa
+4\gamma_{1})(\gamma_{1}^{2}-\gamma_{2}^{2})}
{(\kappa +\gamma_{1})(\kappa +2\gamma_{1})^{2}}\right] \right\}\nonumber \\
&-\frac{2g_{1}^{4}}{\kappa^{2}\Omega^{2}}\left\{1-\frac{\left(\gamma_{1}^{2}-\gamma_{2}^{2}\right)}
{\gamma_{1}\left(\gamma_{1}+\kappa\right)}\right. \nonumber \\
&\left.
+\frac{32g_{1}^{2}}{\kappa^{2}}\left[1-\frac{u\left(\gamma_{1}^{2}-\gamma_{2}^{2}\right)}
{\gamma_{1}\left(\gamma_{1}+\kappa\right)^{2}\left(2\gamma_{1}+\kappa\right)^{2}}\right]\right\},\nonumber \\
\langle{a^{\dag}a}\rangle&=\frac{4g_{1}^{2}}{\kappa^{2}}\left(1 -
\frac{2\left(\gamma_{1}^{2}-\gamma_{2}^{2}\right)}
{\gamma_{1}\left(2\gamma_{1}+\kappa\right)}
+\frac{g_{1}^{2}}{2\Omega^{2}}\left\{1-\frac{\left(\gamma_{1}^{2}-\gamma_{2}^{2}\right)}
{\gamma_{1}\left(\gamma_{1}+\kappa\right)}\right.\right. \nonumber \\
&\left.\left. +\frac{32g_{1}^{2}}{\kappa^{2}}\left[1-\frac{u\left(\gamma_{1}^{2}-\gamma_{2}^{2}\right)}
{\gamma_{1}\left(\gamma_{1}+\kappa\right)^{2}\left(2\gamma_{1}+\kappa\right) ^{2}}\right]\right\}\right)
 ,\label{e22}
\end{align}
where $u=\left(\gamma_{1}+\kappa\right)\left(2\gamma_{1}+\kappa\right)\left(2\gamma_{1}+3\kappa\right)
+\left(4\gamma_{1}+3\kappa\right)\gamma_{2}^{2}$.

Similarly to the dressed-atom inversion, the properties of the expectation values of the cavity field
depend on the parameters $\gamma_{1}$ and $\gamma_{2}$ and thus are also sensitive to the
mode structure of the band gap material. A particular situation of interest is that of
$\gamma_{1}=\gamma_{2}$ which, according to Eq.~(\ref{e21a}) occurs when the dephasing rate
$\gamma_{p}=0$ and the lower Rabi sideband frequency is adjusted to be inside the forbidden
frequency region of the band gap material, $\gamma_{-}=0$. As we see from Eq.~(\ref{e22}),
the cavity field correlation functions are then enhanced due to the combined effect of the band gap
material and the non-secular processes.

\begin{figure}[hbp]
\includegraphics[width=\columnwidth,keepaspectratio,clip]{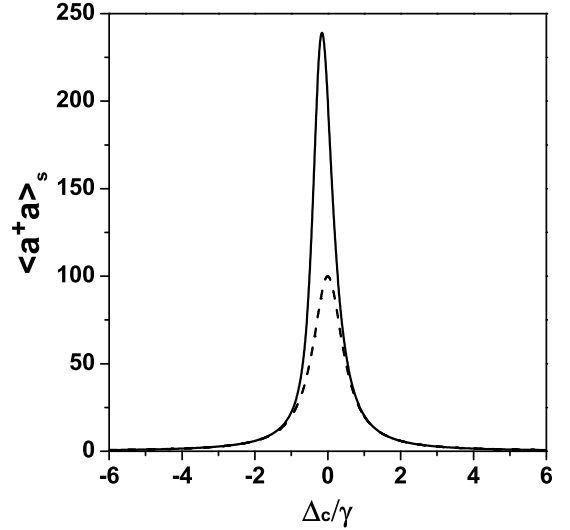}
\caption{The steady-state expectation value of the number of photons
in the cavity field plotted as a function of the detuning
$\Delta_{c}$ in the presence (solid line) and in the absence (dashed
line) of the non-secular terms for $\Omega=100\gamma$,
$\gamma_{+}=10\gamma$, $\kappa=0.1\gamma_{+}$, $g=10\kappa$,
$\Delta_{a}=0$, and $\gamma_{-}=\gamma_{p}=0$. } \label{fig1}
\end{figure}

\ \ The non-secular terms not only enhance the number of photons,
but also lead to a shift of the cavity resonance frequency. This is
shown in Fig.~\ref{fig1}, where we plot the steady-state expectation
value of the number of photons in the cavity field as a function of
the detuning $\Delta_{c}$ in the band gap material configuration
$\gamma_{-}=0$, in which spontaneous emission is forbidden at the
lower Rabi sideband frequency. The graphs are computed from the
exact steady-state solution of Eq.~(\ref{e20}) that is presented in
the Appendix I. It is evident that in the presence of the
non-secular terms the number of photons attains the maximum  value
in the vicinity of the detuning $\Delta_{c}= -0.16\gamma$. The shift
of the maximum towards the negative detuning $\Delta_{c}$ indicates
a pulling of the oscillation frequency of the cavity field away from
the passive cavity frequency $\omega_{c}$ towards the lower Rabi
sideband frequency that is inside the forbidden frequency region of
the band gap material. This is another noteworthy feature of the
non-secular terms in the dressed-atom picture.

\ \ We note here that the values of the parameters used to plot Fig.~\ref{fig1} are 
consistent with the good cavity limit assumed in the derivation of the master equation (\ref{ma1})
and are experimentally realistic. We have discussed the validity of the good cavity limit 
in Sec.~\ref{sec2} for realistic parameters of our system, and have shown that a small cavity damping 
and a large atom cavity coupling $g$ can be readily achieved within the current band-gap material 
technology~\cite{lf,gr}.

\section{Fluctuations of the cavity field}

\ \ We now proceed to investigate the fluctuations of the cavity
field amplitude by analyzing the variances of the conjugate
quadrature phase amplitudes
$X_{+}=ae^{i\theta}+a^{\dag}e^{-i\theta}$ and
$X_{-}=-i(ae^{i\theta}-a^{\dag}e^{-i\theta})$, where $\theta$ is the
quadrature phase. Previous calculations of Florescu {\it et
al.}~\cite{lf} have shown that under the secular approximation and
by appropriate tuning of the dressed-atom transition frequencies to
the band gap material mode transfer structure, the fluctuations of
the cavity field can be reduced down up to the quantum shot noise
limit. We now consider the possibility of using the non-secular
processes to reduce the quantum fluctuations below the quantum
limit. There is a good reason to believe that the non-secular
processes can modify the fluctuations of the cavity field amplitude.
They appear as non-linear processes and it is well known that these
processes are typical sources used to generate squeezed
light~\cite{df04}.

\ \ Let us first examine the normally ordered variance of the
quadrature component $X_{+}$ for the simplified case of
$\Delta_{c}=0$. This analysis provides simple analytical formulas
for the variance and illustrates the role of the non-secular
processes in the behavior of the cavity field. To determine the
variance, we make use Eq.~(\ref{e22}) and find that it can be
expressed as
\begin{equation}
\langle :\left({\Delta{X}_{+}}\right)^{2}:\rangle_{s} = S_{1} + S_{2} ,
\end{equation}
where
\begin{eqnarray}
S_{1} = \frac{8g_{1}^{2}(\gamma_{1}^{2}-\gamma_{2}^{2})}
{\kappa\gamma_{1}^2(\kappa+2\gamma_{1})}(1+ \cos{2\theta}) \label{s1}
\end{eqnarray}
is the contribution of the linear interaction term, and
\begin{align}
S_{2} =& -\frac{2g_{1}^{2}\gamma_{2}\sin{2\theta}}
{\kappa\gamma_{1}\Omega}\left[1
+\frac{32g_{1}^{2}(\gamma_{1}^{2}-\gamma_{2}^{2})(3\gamma_{1}+2\kappa)}
{\kappa\gamma_{1}(\kappa+\gamma_{1})(\kappa+2\gamma_{1})^2}\right ]\nonumber \\
&+\frac{4g_{1}^{4}(1-\cos{2\theta})}{\kappa^{2}\Omega^{2}}
\left\{1 -\frac{(\gamma_{1}^{2}-\gamma_{2}^{2})}
{\gamma_{1}(\kappa+\gamma_{1})}\right. \label{s2}\\
&\left. +\frac{32g_{1}^{2}(\gamma_{1}^{2}-\gamma_{2}^{2})\left[\kappa\gamma_{1}
(\kappa+\gamma_{1})+(5\gamma_{1}+4\kappa)\gamma_{2}^{2}\right]}
{\kappa\gamma_{1}^{2}(\kappa +\gamma_{1})^{2}(\kappa +2\gamma_{1})^{2}}\right\}
\nonumber
\end{align}
is the contribution from the non-secular terms.

\ \ Clearly, $S_{1}\geq 0$ and $S_{2}$ vanishes under the secular
approximation. Note that the term $S_{2}$ is negative for the phase
angle $0<\theta <\pi/2$ and $\gamma_{2}$ positive. Thus, for an
appropriate choice of the parameters, the fluctuations of the cavity
field can be squeezed below the quantum shot noise level. However,
when we examine Eqs.~(\ref{s1}) and (\ref{s2}), we find that the
$S_{1}$ term exceeds the $S_{2}$ term independent of the values of
the parameters involved. This means that  squeezing is possible only
when the term $S_{1}$ is suppressed. Inspection of Eq.~(\ref{s2})
shows that the $S_{1}$ term is suppressed when
$\gamma_{1}=\gamma_{2}$. According to Eq.~(\ref{e21a}), this
conditions can be achieved by suppressing spontaneous emission on
transitions at the lower Rabi frequency $\omega_{L}-2\Omega$. Thus,
the immediate consequence of the fact that $\gamma_{1}=\gamma_{2}$
is that the dominating positive term $S_{1}$ vanishes, leaving the
variance depending solely on the term $S_{2}$ that arises from the
presence of the non-linear processes between the dressed states.
\begin{figure}[hbp]
\includegraphics[width=\columnwidth,keepaspectratio,clip]{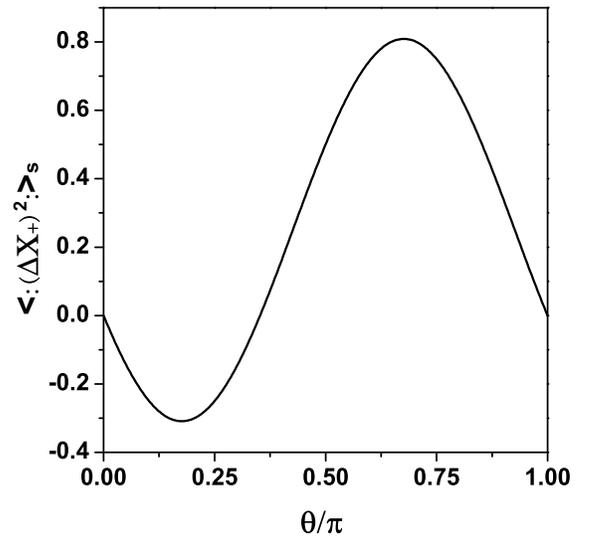}
\caption{The steady-state variance $\langle :(\Delta{X_{+}})^{2}:\rangle_{s}$ 
of the quadrature component of the
cavity field as functions of the quadrature phase~$\theta/\pi$ in a full photonic
band gap and for vanishing detuning $\Delta_{c}=0$. All other
parameters are same as in Fig.~1.} \label{fig2}
\end{figure}
Note the dependence of the "squeezer", the $S_{2}$ term of the variance, on the ratio
$\gamma_{2}/\gamma_{1}$ that according to Eq.~(\ref{e21}) determines the inversion between
the dressed states. Apparently, for phase angles $0<\theta <\pi/2$, squeezing is obtained only
when $\gamma_{2}>0$, i.e. when there is no inversion between the dressed states and vice versa, for
phase angles $\pi/2<\theta <\pi$, squeezing is obtained for a negative $\gamma_{2}$, i.e. when there
is an inversion between the dressed states.
Moreover, the optimal condition for squeezing is to maintain
$\gamma_{1}=\gamma_{2}$. These analysis indicate that a photonic band gap material can be employed to generate a squeezed field from a single-atom laser.

\ \ Figure~\ref{fig2} shows the normally ordered variance $\langle
:({\Delta{X}_{+}})^{2}:\rangle_{s}$ as a function of the quadrature
phase angle $\theta$. As the phase change, the variance oscillates
from negative to positive values. Negative values mean squeezing of
the fluctuations of the quadrature phase amplitude below the quantum
shot noise level. The maximum negative value of the variance,
corresponding to the maximum squeezing occurs at $\theta \approx
0.55$. However, the maximum value of squeezing obtained for
$\Delta_{c}=0$ might not be the greatest value possible due to the
shift of the cavity resonance. Actually, it appears for a detuning
slightly shifted from the cavity resonance. This is shown in
Fig.~\ref{fig3}, where we plot the variance as a function of the
scaled Rabi frequency $\epsilon/\Delta_{a}$, for
$\gamma_{1}=\gamma_{2}$ and different detunings $\Delta_{c}$. We see
that the variance is negative for all values of the Rabi frequency,
thus indicating that the cavity field can be in a squeezed state
irrespective of the strength of the laser field. As the Rabi
frequency increases, the variance decreases and then saturates to a
constant value. The saturation level decreases with increasing
negative detuning $\Delta_{c}$ and attains the maximal negative
value, corresponding to the maximum squeezing, for
$\Delta_{c}=-0.26\gamma$.

\begin{figure}[hbp]
\includegraphics[width=\columnwidth,keepaspectratio,clip]{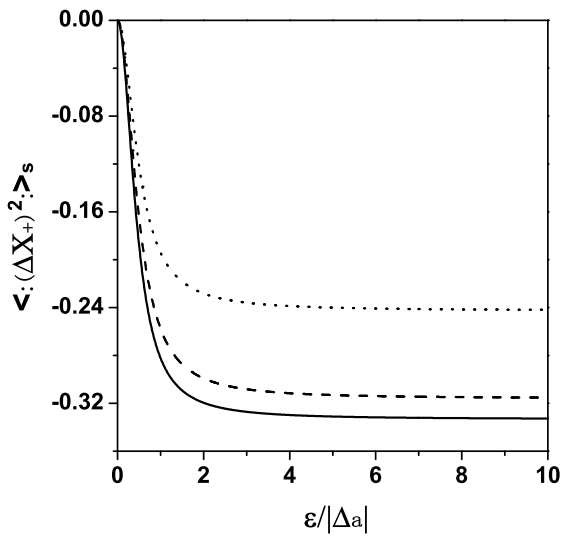}
\caption{The normally ordered variance $\langle
:(\Delta{X_{+}})^{2}:\rangle_{s}$ as a function of the scaled Rabi
frequency $\varepsilon/\Delta_{a}$ in the band gap material
configuration of $\gamma_{-}=0$ for $\theta=0.8$,
$\Omega=100\gamma$, $\gamma_{+}=10\gamma$, $\kappa=0.1\gamma_{+}$,
$g=10\kappa$, $\gamma_{p} =0$ and different detunings  $\Delta_{c}$:
$\Delta_{c}=0$ (dotted line), $\Delta_{c}=-0.16\gamma$ (dashed line)
and $\Delta_{c}=-0.26\gamma$ (solid line).} \label{fig3}
\end{figure}

\ \ We may conclude this section that the filtering properties of
the band gap material and the treatment of the interaction between
the cavity field and the dressed atom beyond the secular
approximation produces important modifications in the variance of
the quadrature field amplitudes, greatly reducing the fluctuations
below the quantum limit.

\section{Spectrum of the cavity field}

\ \ In this section we investigate the spectrum of the cavity field
and point out one more difference between the secular and
non-secular single-atom lasers. It has been shown by 
Florescu {\it et al.}~\cite{lf} that in the secular approximation, the spectral
line of the cavity field can be narrowed to the quantum limit. Here,
we will illustrate that the linewidth of the spectrum can be further
reduced that it may go below the quantum limit if the non-secular
processes are included into the dynamics of the system.

\ \ The stedy-state spectrum of the cavity field is defined as the
Fourier transform of the two-time correlation function
$\langle{a}^{\dag}(t)a\rangle_{s}$ of the cavity field operators. In
the interaction picture, the spectrum takes the form
\begin{equation}
S(\omega)=2\mathrm{Re}\int_{0}^{\infty}dt{e}^{i(\omega-\omega_{L})t}
\langle{a}^{\dag}(t){a}\rangle_{s} ,
\end{equation}
where  $\omega$ is the spectral frequency, the subscript $s$ represents the average over the steady-state
values of the field operators, and the operators without time argument refer to their steady-state values.

\ \ We may write the cavity field operators in the form
$a(t)=\langle{a}(t)\rangle+\delta{a(t)}$, where $\delta{a(t)}$ are
the fluctuations operators that describe fluctuations of the cavity
field about its average value. In this case,  the spectrum can be
decomposed into a sum of two terms
\begin{equation}
S(\omega)=S_{el}(\omega)+S_{in}(\omega) ,
\end{equation}
where
\begin{equation}
S_{el}(\omega)=2\pi\langle{a^{\dag}}\rangle_{s}\langle{a}\rangle_{s}\delta(\omega-\omega_{L})
\end{equation}
is the elastic component of the spectrum, and
\begin{eqnarray}
S_{in}(\omega)=2\mathrm{Re}\int_{0}^{\infty}dt{e}^{i(\omega-\omega_{L})t}
\langle a^{\dagger}(t),a\rangle_{s} \label{e32}
\end{eqnarray}
is the inelastic component with
\begin{eqnarray}
\langle a^{\dagger}(t),a\rangle_{s}
= \langle{a}^{\dag}(t){a}\rangle_{s} -\langle{a}^{\dag}(t)\rangle_{s}\langle{a}\rangle_{s} .
\end{eqnarray}

\begin{figure}[hbp]
\includegraphics[width=\columnwidth,keepaspectratio,clip]{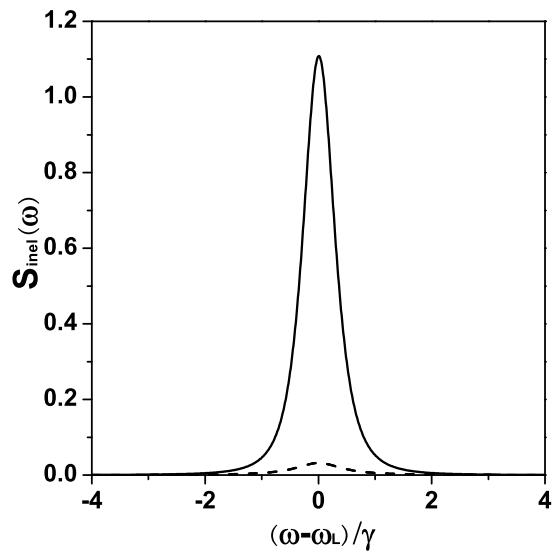}
\caption{The incoherent part of the spectrum of the cavity field as
a function of the frequency $(\omega-\omega_{L})/\gamma$ calculated with
(solid line) and without (dashed line) the non-secular terms for
$\gamma_{-}/\gamma_{+}=0.0001$ and $\Delta_{c}=0$. All other parameters
are same as in Fig.~1.} \label{sp1}
\end{figure}

\begin{figure}[hbp]
\includegraphics[width=\columnwidth,keepaspectratio,clip]{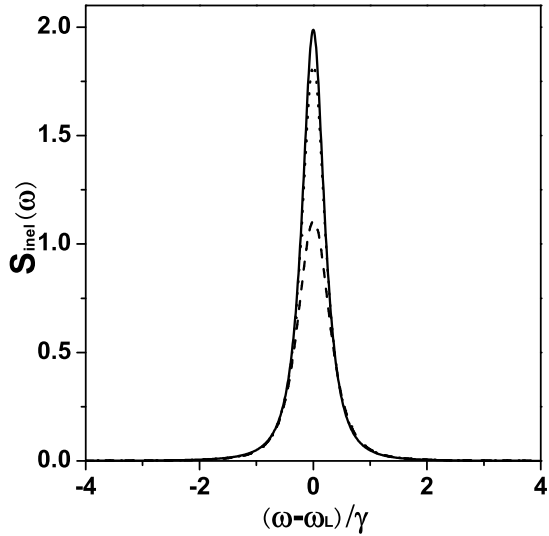}
\caption{The incoherent part of the spectrum of the cavity field as
a function of the frequency $(\omega-\omega_{L})/\gamma$ 
for different detunings $\Delta_{c}$: $\Delta_{c}=0$
(dashed line), $\Delta_{c}=-0.16$ (dotted line) and
$\Delta_{c}=-0.26$ (solid line). All other parameters are same as in
Fig.~1.} \label{sp2}
\end{figure}

In order to calculate the two-time correlation function appearing in Eq.~(\ref{e32}),
we make use of the quantum regression theorem~\cite{lax,hj} from which it is well known that
the two-time correlation functions obey the same equations of motion for $t>0$ as the corresponding one-time correlation functions. Therefore, we may used Eqs.~(\ref{e19}) and (\ref{e20}) to compute the spectrum. The spectrum depends, of course, on the state of the system which we take to be the stationary state.

\ \ We will compute numerically the spectra using the equations of
motion (\ref{e19}) and (\ref{e20}). However, in order to gain
insight into the physics involved, we will present an analytical
solution for the case of $\Delta_{c}=0$.

\ \ For purposes of numerical computation, it is convenient to take
the Laplace transform of the equations of motion for the correlation
functions that transforms them into a set of algebraic equations,
and evaluate the spectrum from
\begin{align}
S_{in}(\omega)&= 2{\rm Re}[\langle a^{\dagger}(p),a\rangle_{s} ]
|_{p=i(\omega-\omega_{L})} ,\label{ins}
\end{align}
where $p$ is a complex Laplace transform parameter. The exact solution for the Laplace transform
of the correlation function with $\Delta_{c}=0$ is given in the Appendix II. We use this exact solution to evaluate the spectrum in the presence of the non-secular processes. We compare the spectra with those obtained under the secular approximation where the non-secular processes are ignored. The results are show in Figs.~\ref{sp1} and \ref{sp2}.

\ \ Let us first compare the spectra obtained with the non-secular
processes with that obtained under the secular approximation.
Figure~\ref{sp1} shows the incoherent part of the spectrum of the
cavity field in the band gap material configuration with
$\gamma_{-}=0.0001\gamma_{+}$. The spectrum consists of a single
line centered at frequency $\omega=\omega_{L}$ whose the profile
depends on whether the non-secular processes are or are not
included. The shape of the line is not exactly a Lorentzian when the
non-secular processes are included, but becomes a Lorentzian when
these processes are ignored. One may notice a significant linewidth
narrowing below the natural width when the non-secular processes are
included.

\ \ Figure~\ref{sp2} illustrates the effect of the cavity detuning
on the spectrum. We see that the maximum narrowing of the spectral
line actually occurs not at $\Delta_{c}=0$ but for the detuning
$\Delta_{c}=-0.26\gamma$ that, according to Fig.~\ref{fig3},
corresponds to the maximum squeezing in one of the quadrature phase
components of the cavity field. This indicates that the spectrum may
be directly affected by squeezing in the cavity field.

\ \ In order to gain insight into the source of the narrowing of the
spectral line, we take the limit of $\Delta_{c}=\Delta_{a}=0$,
$\gamma_{-}=0$ and derive from the general solution, Eq.~(\ref{B1}),
a simple analytical formula for the spectrum
\begin{eqnarray}
S_{in}(\omega)
=\frac{16\kappa{g}_{1}^{4}}{\Omega^{2}\left[\kappa^2+4(\omega-\omega_L)^2\right]^2} .
\end{eqnarray}
This result is in marked contrast to that obtained under the secular
approximation~\cite{lf}. In the case of vanishing mode density on
the lower Mollow sideband ($\gamma_{-}=0$, which corresponds to a
full photonic band gap) and no dipolar dephasing ($\gamma_{p}=0$),
the spectrum of the cavity field consists only of the elastic
component under the secular approximation~\cite{lf}. When the non-secular processes
are included, the incoherent part of the spectrum is present and
shows an interesting property that instead of being a simple
Lorentzian, it is instead in the form of a squared Lorentzian.

\begin{figure}[hbp]
\includegraphics[width=\columnwidth,keepaspectratio,clip]{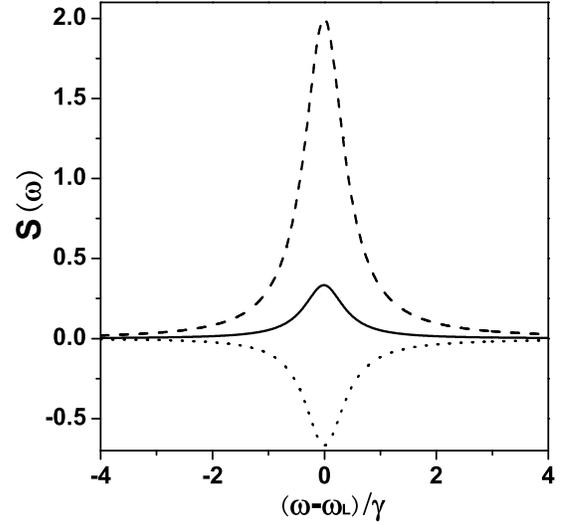}
\caption{The incoherent part (solid line) of the cavity field
spectrum together with the squeezing spectra $X_{+}(\omega)$ (dotted
line) and $X_{-}(\omega)$ (dashed line) plotted as a function of
$(\omega -\omega_{L})/\gamma$  for
$\Delta_{c}=-0.26\gamma$ and $\theta=0.8$. All other parameters are
same as in Fig.~5.} \label{sp3}
\end{figure}

\ \ It is well known that a squared Lorentzian can be decomposed
into a difference between two Lorentzians ~\cite{rc88,fs97,rs01}.
The immediate consequence
of one of the Lorentzians being negative is to produce a spectrum
which fell off as $\omega^{-4}$ in the wings, rather than the
$\omega^{-2}$, which would result if both Lorentzians were positive.
It has been shown  that the negative weight of one of the
Lorentzians can be related to squeezing in the field. Thus, the
narrowing of the spectral line, seen in Figs.~\ref{sp1} and
\ref{sp2} can be related to squeezing produced in the non-linear
interaction of the cavity field with the dressed atom. Physically,
this effect can be best understood and explained in terms of the
so-called squeezing spectrum. The incoherent spectrum can be
decomposed into the sum of two phase dependent squeezing spectra
\begin{eqnarray}
S_{in}(\omega) = \frac{1}{4}\left[ X_{+}(\omega) +X_{-}(\omega)\right] ,
\end{eqnarray}
where
\begin{eqnarray}
X_{+}(\omega) &=& \int_{-\infty}^{\infty} dt{e}^{i(\omega-\omega_{L})t}\langle :X_{+}(t),X_{+}:\rangle ,\nonumber \\
X_{-}(\omega) &=&  \int_{-\infty}^{\infty} dt{e}^{i(\omega-\omega_{L})t}\langle : X_{-}(t),X_{-}:\rangle ,
\end{eqnarray}
are the squeezing spectra of the normally ordered quadrature phase components of the cavity field amplitude.
Negative values in the squeezing spectra indicate squeezed fluctuations of the cavity field.

\ \ Figure~\ref{sp3} shows the incoherent spectrum together with the
squeezing spectra calculated for the same parameters as in
Fig.~\ref{sp2}, but with $\Delta_{c}=-0.26\gamma$ corresponding to
the maximum squeezing in the $X_{+}$ quadrature components of the
cavity field. We see that the squeezing spectrum $X_{+}(\omega)$ is
negative over the whole range of frequencies. The fact that
$X_{+}(\omega)$ is negative indicates that the $X_{+}$ quadrature
component, that is obtained by an integration of the squeezing
spectrum $X_{+}(\omega)$ over the frequency $\omega$, is squeezed.
Thus, we may conclude that the spectral linewidth narrowing is due
to the common effect of squeezing and the frequency dependent
reservoir formed by the photonic band gap material.

\section{Conclusions}
\ \ We have examined nonclassical and spectral features of a
single-atom laser without the use of the secular approximation for
the interaction of the cavity field with a dressed-atom system
located inside a photonic band gap material. The material appears as
a frequency dependent reservoir for the dressed atom. The inclusion
of the non-secular terms leads to an effective double coupling of
the cavity field to the dressed states of the system. We have found
that by a suitable matching of the dressed-atom transition
frequencies to the guiding frequency of the band gap material, it is
possible to account for squeezing and narrowing of the spectral line
of the cavity field below the quantum shot noise limit. These
features are achieved by a combination of a strongly
frequency-dependent reservoir and the non-secular processes. The
non-linear processes are absent under the secular approximation and
present a newly encountered features of a single-atom laser.

\section{Acknowledgments}
The authors acknowledge financial support from the National Natural Science
Foundation of China (Grant No. 10674052), the Ministry of Education
under project NCET (grant no NCET-06-0671) and the National Basic
Research Project of China (grant no 2005 CB724508). ZF would like to thank
the Huazhong Normal University for hospitality.

\section*{APPENDIX I}

In this appendix, we present the exact steady-state solution of the equations of motion (\ref{e19})
and (\ref{e20}) for the case of $\Delta_{c}=0$ that includes all non-secular terms up to the
order $g^{4}/\Omega^{2}$.
\addtocounter{equation}{1}
\begin{align*}
\langle{a}\rangle_{s}&=-2g_{1}[2\alpha{F_{1}}+\gamma_{2}\kappa(2\gamma_{1}+\kappa)^{2}]/K_{1},\tag{\theequation.2}\\
\langle{a^{2}}\rangle_{s}&=\{[-2(\alpha-\alpha')F_{2}+8g_{1}^{2}\alpha{F_{3}}]F_{4} \nonumber\\
&  +4g_{1}^{2}(1-\frac{4\alpha}{\gamma_{2}})F_{5}\}/K_{2}-(\alpha-\alpha')/(2\gamma_{2}),\tag{\theequation.3}\\
\langle{a^{\dag}a}\rangle &=2[(\alpha-\alpha')^{2}F_{2}-4g_{1}^{2}\alpha(\alpha-\alpha')F_{3} \\
&+2g_{1}^{2}F_{5}]/K_{2},\tag{\theequation.4}\\
\langle{R_{3}a^{\dag}a}\rangle_s &=\frac{1}{(\kappa+\gamma_{1})K_{2}\gamma_{2}}
[-2(\alpha-\alpha')^{2}x_{0}F_{2} \\
& +8g_{1}^{2}\alpha(\alpha-\alpha')x_{0}F_{3} -4g_{1}^{2}(\gamma_{2}^{2}
+4\alpha(\alpha-\alpha'))F_{5} \\ &-4g_{1}^{2}\kappa\gamma_{2}^{2}((\kappa+2\gamma_{1})^{2}
-4\alpha\alpha')K_{3}\\
&-(\alpha-\alpha')^{2}K_{2}/2],\tag{\theequation.5}\\
\langle{R_{3}a}\rangle_s &=\frac{2g_{1}}{K_{1}}[-4(\kappa\gamma_{1} 
+2\gamma_{2}\alpha)\alpha\alpha' \\
&+2\gamma_{2}(\kappa^{2}+4\gamma_{2}^{2} +4\kappa\gamma_{1})\alpha \\
&+\kappa(\kappa\gamma_{1}+2\gamma_{2}^{2})(2\gamma_{1}+\kappa)],\tag{\theequation.6}\\
\langle{R_{3}a^{2}}\rangle_s&=\frac{2}{(\kappa+\gamma_{1})}\{[(\alpha-\alpha')x_{1}F_{2}-4g_{1}^{2}\alpha{x_{1}}F_{3} \\
&+2g_{1}^{2}x_{2}F_{5}]/K_{2}+x_{3}/4\} ,\tag{\theequation.7}\\
\end{align*}
where
\begin{align*}
F_{1}&=-4\alpha\alpha^{\prime}\gamma_{1} -2\gamma_{2}\kappa\alpha'+(2\gamma_{1}
+\kappa)(\kappa\gamma_{1}+2\gamma_{2}^{2}),\\
F_{2}&=16(\alpha\alpha')^{3}\gamma_{1} \\
&-4[3\gamma_{1}(\kappa+\gamma_{1})^{2}+2\gamma_{1}^{3}+(\gamma_{2}^{2}-\gamma_{1}^{2})(\kappa+5\gamma_{1})](\alpha\alpha')^{2}\\
&+\{\gamma_{1}\kappa^{2}(2\gamma_{1}+\kappa)^{2}+2(\kappa+\gamma_{1})(\kappa\gamma_{1}+\gamma_{2}^{2})[(\kappa+\gamma_{1})^{2} \\
&+2\gamma_{2}^{2}-\gamma_{1}^{2}]\}\alpha\alpha'\\
&-\kappa^{2}(\kappa+\gamma_{1})(\kappa\gamma_{1}+\gamma_{2}^{2})(2\gamma_{1}+\kappa)^{2}/4,\\
F_{3}&=32(\alpha\alpha')^{2}\gamma_{1} \\
&-8[(\kappa\gamma_{1}+2\gamma_{2}^{2})(2\gamma_{1}+\kappa)+(\kappa+\gamma_{1})(\kappa\gamma_{1}+\gamma_{2}^{2})]\alpha\alpha'\\
&+(2\gamma_{1}+\kappa)(\gamma_{1}\kappa+2\gamma_{2}^{2})(2\kappa^{2}+2\kappa\gamma_{1}+\gamma_{2}^{2})\\
&+\gamma_{2}^{2}(2\gamma_{1}+\kappa)^{2}(\gamma_{1}+2\kappa)-4\gamma_{2}^{2}(\gamma_{1}+\kappa)(\gamma_{1}^{2}-\gamma_{2}^{2}),\\
F_{4}&=[\kappa(\kappa+\gamma_{1})-4\alpha\alpha']/\gamma_{2}+\alpha+\alpha',\\
F_{5}&= -64(\alpha\alpha')^{3}\gamma_{1} \\
&+16[(3\kappa\gamma_{1}+2\gamma_{2}^{2})(\kappa+\gamma_{1})+\gamma_{1}(3\gamma_{2}^{2}+\kappa\gamma_{1})](\alpha\alpha')^{2}\\
&-4[(\kappa\gamma_{1}+2\gamma_{2}^{2})(2\gamma_{1}+\kappa)(2\kappa\gamma_{1}+2\kappa^{2}+\gamma_{2}^{2}) \\
&+\gamma_{1}\kappa^{2}(\kappa+\gamma_{1})^{2}]\alpha\alpha'\\
&+\kappa^{2}(2\gamma_{1}+\kappa)(\kappa\gamma_{1}+2\gamma_{2}^{2})(\kappa+\gamma_{1})^{2} ,
\end{align*}
and
\begin{align*}
K_{1}&= \{ [4\alpha\alpha' -\kappa(2\gamma_{1}+\kappa)]^{2} -16\gamma_{2}^{2}\alpha\alpha'\}\gamma_{1},\\
K_{2}&=\{256(\alpha\alpha')^{4}-64(4\kappa^{2}+6\kappa\gamma_{1}+5\gamma_{2}^{2})(\alpha\alpha')^{3} \\
&+16[2(\kappa^{2}+2\kappa\gamma_{1}+2\gamma_{2}^{2})(2\kappa^{2}+2\kappa\gamma_{1}+\gamma_{2}^{2})\\
&+\kappa^{2}((2\gamma_{1}+\kappa)^{2}+(\kappa+\gamma_{1})^{2})](\alpha\alpha')^{2} \\
&-4\kappa^{2}[2(\kappa^{2}+2\kappa\gamma_{1}+2\gamma_{2}^{2})(\gamma_{1}+\kappa)^{2}\\
&+(2\kappa^{2}+2\kappa\gamma_{1}+\gamma_{2}^{2})(\kappa+2\gamma_{1})^{2}]\alpha\alpha' \\
&+\kappa^{4}(\kappa+\gamma_{1})^{2}(\kappa+2\gamma_{1})^{2}\}\gamma_{1},\\
K_{3}&=  [4\alpha\alpha' -\kappa(\gamma_{1}+\kappa)]^{2} -4\gamma_{2}^{2}\alpha\alpha' ,
\end{align*}
with
\begin{align*}
x_{0}&=\gamma_{2}^{2}+\kappa(\kappa+\gamma_{1})-4\alpha\alpha',\\
x_{1}&=x_{0}(\alpha+\alpha')/\gamma_{2}+\kappa(\kappa+\gamma_{1}),\\
x_{2}&=-\gamma_{2}+4\alpha(\alpha+\alpha')/\gamma_{2},\\
x_{3}&=(\alpha^{2}-\alpha'^{2})/\gamma_{2} \\
&-8g_{1}^{2}\kappa\gamma_{2}[(\kappa+2\gamma_{1})^{2}-4\alpha\alpha']K_{3}/K_{2} ,\\
\alpha&=i\frac{g^{2}}{2\Omega},\\
\alpha^{\prime}&=i\frac{g^{2}(s^2-c^2)^{2}}{2\Omega}   .
\end{align*}

\section*{APPENDIX II}

In this appendix, we present the Laplace transform of the two-time
correlation function of the cavity field operators appearing in the
expression for the incoherent part of the spectrum. We have applied
the quantum regression theorem to Eqs.~(\ref{e19}) and (\ref{e20})
and calculated the Laplace transform with the initial condition
given by the stationary solutions listed in the Appendix I for the
case of the cavity frequency tuned on resonance with the central
frequency of the dressed-atom system, $\Delta_{c}=0$.
\begin{eqnarray}
\langle a^{\dag}(p), a\rangle_{s} = \frac{m_{4}p^{4}+m_{3}p^{3}+m_{2}p^{2}+m_{1}p+m_{0}}
{2(p+\gamma_{1})D(p)} ,\label{B1}
\end{eqnarray}
where
\begin{eqnarray}
D(p) = \alpha\alpha' \gamma_{2}^{2}
-[(p+\frac{\kappa}{2})(p+\frac{\kappa}{2}+\gamma_{1})-\alpha\alpha']^{2} ,
\end{eqnarray}
and
\begin{align*}
m_{0}&=[2(g_{1}\langle{R_{3}a}\rangle_{s}+\gamma_{1}\langle{a^{\dag}a}\rangle_{s})(\gamma_{1}+\kappa/2)
+((\alpha-\alpha')\langle{R_{3}a^{2}}\rangle_{s} \\
&+(\alpha+\alpha')\langle{R_{3}a^{\dag}a}\rangle_{s})\gamma_{1}]
[-\kappa(\gamma_{1}+\kappa/2)/2+\alpha\alpha'] \\
&+2\gamma_{1}[\kappa(\gamma_{1}+\kappa/2)-2\alpha\alpha'](\gamma_{1}+\kappa)\langle{a}\rangle_{s}\langle{a^{\dag}}\rangle_{s}\\
&+\gamma_{2}(\gamma_{1}+\kappa/2)[\gamma_{1}((\alpha+\alpha')\langle{a^{\dag}a}\rangle_{s}+(\alpha-\alpha')\langle{a^{2}}\rangle_{s}) \\
&+2g_{1}\alpha\langle{R_{3}a}\rangle_{s}]+2\gamma_{1}\gamma_{2}\alpha\alpha'\langle{R_{3}a^{\dag}a}\rangle_{s}\\
&+2g_{1}[\gamma_{1}(\gamma_{2}-\alpha)(\gamma_{1}+\kappa)+\gamma_{2}\kappa(\gamma_{1}^{2}-\kappa^{2}/4+\gamma_{1}\kappa/2 \\
&+\gamma_{2}\alpha+\alpha\alpha')/(2\gamma_{1})]\langle{a}\rangle_{s},\\
m_{1}&=[2g_{1}\langle{R_{3}a}\rangle_{s}+(4\gamma_{1}+\kappa)\langle{a^{\dag}a}\rangle_{s}
+(\alpha-\alpha')\langle{R_{3}a^{2}}\rangle_{s} \\
&+(\alpha+\alpha')\langle{R_{3}a^{\dag}a}\rangle_{s}][-\kappa(\gamma_{1}+\kappa/2)/2+\alpha\alpha']\\
&-(\gamma_{1}+\kappa)[2(g_{1}\langle{R_{3}a}\rangle_{s}+\gamma_{1}\langle{a^{\dag}a}\rangle_{s})(\gamma_{1}+\kappa/2)\\
&+\gamma_{1}(\alpha-\alpha')\langle{R_{3}a^{2}}\rangle_{s}+\gamma_{1}(\alpha+\alpha')\langle{R_{3}a^{\dag}a}\rangle_{s}]\\
&+2[\gamma_{1}((\gamma_{1}+\kappa)^{2}+\kappa(\gamma_{1}+\kappa/2)-2\alpha\alpha')+(\kappa(\gamma_{1}+\kappa/2)\\
&-2\alpha\alpha')(\gamma_{1}+\kappa)]\langle{a}\rangle_{s}\langle{a^{\dag}}\rangle_{s}\\
&+[(2\gamma_{1}+\kappa/2)((\alpha+\alpha')\langle{a^{\dag}a}\rangle_{s}
+(\alpha-\alpha')\langle{a^{2}}\rangle_{s}) \\
&+2g_{1}\alpha\langle{R_{3}a}\rangle_{s}+2\alpha\alpha'\langle{R_{3}a^{\dag}a}\rangle_{s}]\gamma_{2}\\
&+2g_{1}[(\gamma_{2}-\alpha)(2\gamma_{1}+\kappa)+\gamma_{2}\kappa/2]\langle{a}\rangle_{s},\\
m_{2}&=-[(\alpha-\alpha')\langle{R_{3}a^{2}}\rangle_{s}+(\alpha+\alpha')\langle{R_{3}a^{\dag}a}\rangle_{s}]
(\kappa+2\gamma_{1}) \\
&-g_{1}(4\gamma_{1}+3\kappa)\langle{R_{3}a}\rangle_{s}\\
&+[-\gamma_{1}(4\gamma_{1}+3\kappa)-(3\kappa/2+\gamma_{1})(\kappa+2\gamma_{1}) \\
&+\gamma_{2}(\alpha+\alpha')+2\alpha\alpha']\langle{a^{\dag}a}\rangle_{s}
+\gamma_{2}(\alpha-\alpha')\langle{a^{2}}\rangle_{s} \\
&+2[(\gamma_{1}+\kappa)^{2}+2\gamma_{1}(\gamma_{1}+\kappa)+\kappa(\gamma_{1}+\kappa/2) \\
&-2\alpha\alpha']\langle{a}\rangle_{s}\langle{a^{\dag}}\rangle_{s} +2g_{1}(\gamma_{2}-\alpha)\langle{a}\rangle_{s},\\
m_{3}&=-(\alpha-\alpha')\langle{R_{3}a^{2}}\rangle_{s}-(\alpha+\alpha')\langle{R_{3}a^{\dag}a}\rangle_{s}
-2g_{1}\langle{R_{3}a}\rangle_{s} \\
&-3(2\gamma_{1}+\kappa)\langle{a^{\dag}a}\rangle_{s} +2(3\gamma_{1}
+2\kappa)\langle{a}\rangle_{s}\langle{a^{\dag}}\rangle_{s},\\
m_{4}&=2(\langle{a}\rangle_{s}\langle{a^{\dag}}\rangle_{s}-\langle{a^{\dag}a}\rangle_{s}).
\end{align*}

\end{document}